\documentclass[11pt,a4paper]{article}
\usepackage{amsmath,amssymb,amsfonts,a4wide,graphicx,bm,times}
\usepackage{cite}
\makeatletter \let\old@startsection=\@startsection
\renewcommand{\@startsection}[6]
{\old@startsection{#1}{#2}{#3}{#4}{#5}{#6\mathversion{bold}}}
\makeatother

\makeatletter

\@addtoreset{equation}{section}
\makeatother
\let\refOld\ref
\renewcommand{\ref}[1]{(\refOld{#1})}

 \def\d{\delta}

 \def\p{\partial}
 
 \def\a{\alpha}
 \def\b{\beta}
 \def\g{\gamma}
 \def\d{\delta}
 \def\e{\epsilon}

 \def\l{\lambda}

 \def\G{\Gamma}

 \def\o{\omega }
 
\def\hf{\dfrac{1}{2}}

\def\implies{\quad\Rightarrow\quad}

\def\ksc{k_\text{sc}}
\def\vphi{\varphi}

\def\vphiB{\varphi^{(B)}}

\def\CF{\mathcal{F}}

\begin{document}
\begin{titlepage}
\renewcommand{\thefootnote}{\fnsymbol{footnote}}

\vspace*{1cm}
    \begin{Large}
       \begin{center}
         {A note on the integral equation for the Wilson loop in\\ $\mathcal{N} = 2$ $D=4$ superconformal Yang-Mills theory}
       \end{center}
    \end{Large}
\vspace{0.7cm}

\begin{center}
Jean-Emile B{\sc ourgine}\footnote
            {
e-mail address : 
jebourgine@sogang.ac.kr}\\
      
\vspace{0.7cm}                    
{\it Center for Quantum Spacetime (CQUeST)
}\\
{\it Sogang University, Seoul 121-742, Korea}
\end{center}

\vspace{0.7cm}

\begin{abstract}
\noindent

We propose an alternative method to study the saddle point equation in the strong coupling limit for the Wilson loop in $\mathcal{N} = 2$ $D=4$ super Yang-Mills with an $SU(N)$ gauge group and $2N$ hypermultiplets. This method is based on an approximation of the integral equation kernel which allows to solve the simplified problem exactly. To determine the accuracy of this approximation, we compare our results to those obtained recently by Passerini and Zarembo. Although less precise, this simpler approach provides an explicit expression for the density of eigenvalues that is used to derive the planar free energy.

\end{abstract}
\vfill

\end{titlepage}
\vfil\eject

\setcounter{footnote}{0}

\section{Introduction}
One way to obtain non-perturbative results in supersymmetric gauge theories is to use the localization technique to rewrite some of the observables, such as the partition function or the Wilson loops, as multivariable integrals \cite{Pestun2007}. These integrals are reminiscent of the eigenvalue integrals that appear in  matrix models \cite{DiFrancesco1995}. The number $N$ of integration variables is associated to the size of a gauge group and the large $N$ expansion interpreted as a 't Hooft expansion. To compute the integrals order by order in this $1/N^2$ topological expansion, various techniques can be borrowed from matrix models. In particular, the planar order can be obtained using a saddle point approximation: the integration variables, or `eigenvalues', condense on a continuous support and the saddle point equation takes the form of an integral equation for their density. This approach has proved to be very efficient to study three dimensional supersymmetric gauge theories, such as Chern-Simons or ABJM (see \cite{Marino2011} and references therein).

In four dimensions, the Wilson loop and free energy of the $\mathcal{N}=4$ super Yang-Mills theory are described by the standard Hermitian matrix model with Gaussian potential. This was first conjectured in \cite{Erickson2000,Drukker2000} after a perturbative analysis, and later proved by Pestun using localization \cite{Pestun2007}. In this simple case, the eigenvalue integrals can be performed exactly, leading to an expression of the Wilson loop valid at all order in $N$ and for all values of the 't Hooft coupling $\l=Ng^2$, where $g$ is the Yang-Mills coupling. This gauge theory is known to be dual to the type IIB string theory in $AdS_5\times S_5$ background \cite{Maldacena1997}. At large $\l$, the Wilson loop is characterized by an exponential behavior in $\sqrt{\l}$ at the planar order, which agrees with the supergravity result \cite{Erickson2000}. The subleading orders were also successfully compared to the dual string amplitude in \cite{Drukker2000}.

The $\mathcal{N}=2$ $4d$ super Yang-Mills theory with an $SU(N)$ gauge group and $2N$ fundamental hypermultiplets, also referred as superconformal QCD (SCQCD), is more intriguing. Although several proposal were made \cite{Gaiotto2009,Gadde2009,ReidEdwards2010,Colgain2011}, the dual string background has yet to be identified. On the gauge theory side, localization can still be used to express the partition function and Wilson loop as integrals over real eigenvalues. But the Van der Monde determinant of the Hermitian matrix model is now replaced by a product involving G-Barnes functions arising from the one-loop contribution \cite{Pestun2007}.\footnote{The instanton corrections are expected to give exponentially decreasing corrections at large $N$ \cite{Passerini2011} and will be neglected in this note.} This model was first proposed in \cite{Rey2010}, where a scaling argument has been employed to predict a very unusual power growth of the planar Wilson loop at large $\l$. This fact was later confirmed by Passerini and Zarembo who used the saddle point technique to derive the expression of the Wilson loop at strong coupling,
\begin{equation}\label{Wilson_PZ}
W(\l)=A\left(\dfrac{\l}{\sqrt{\log\l}}\right)^{3},\quad A\simeq 9.47\times 10^{-5}.
\end{equation}
We briefly summarize their computation below.

In the saddle point approximation, the planar Wilson loop is written as an integral over the density of eigenvalues,
\begin{equation}\label{def_Wilson}
W(\l)=\int_\G{e^{2\pi x}\rho(x)dx},
\end{equation}
which satisfies a singular integral equation, the saddle point equation,
\begin{equation}\label{equ_orig}
\int_\G{\dfrac{\rho(y)dy}{x-y}}-\int_\G{k(x-y)\rho(y)dy}=\dfrac{8\pi^2}{\l}x-k(x).
\end{equation}
The RHS of this equation is the derivative of the matrix potential, but it will simply be referred here as the potential of the integral equation. The kernel of the equation \ref{equ_orig} is a linear combination of the usual Cauchy kernel and the function $k$. In the $\mathcal{N}=4$ theory, the function $k$ doesn't appear and we recover the saddle point equation of the Gaussian Hermitian matrix model. The solution is then given by the Wigner semi-circle law, and the endpoints of the compact support $\G$ scale as $\sqrt{\l}$. This square root dependence leads to the Wilson loop exponential behavior at large $\l$. In SCQCD, the function $k$ is the logarithmic derivative of the G-Barnes functions that replace the Van der Monde determinant, it can be expressed as a sum of two digamma functions,
\begin{equation}\label{k_def}
k(x)=x(\psi(ix)+\psi(-ix)+2\g),\quad \psi(1)=-\g.
\end{equation}
In the weak coupling limit $\l\to0$, the function $k$ can be neglected and we recover the $\mathcal{N}=4$ solution. Conversely, at strong coupling this function dominates the Cauchy kernel, leading to an unusual behavior of the density. To investigate the saddle point equation \ref{equ_orig}, the authors of  \cite{Passerini2011} employed the Wiener-Hopf method, a refinement of the Fourier transform for integral equations with compact support. They derived an approximate density in the region $x>0$ which was expressed in the Fourier space. From the unit norm condition, and using the parity of the density, they deduced a logarithmic dependence of the support endpoints in $\l$. This exotic behavior of the support gives rise to the algebraic expression \ref{Wilson_PZ} of the Wilson loop in $\l$ and $\log\l$ at strong coupling. Their method and results were consistently checked by numerical simulations.\\

A possible approach to uncover the string background of SCQCD is to consider $\mathcal{N}=2$ super Yang-Mills with $SU(N)\times SU(N)$ gauge group and bifundamental hypermultiplets. This theory depends on two 't Hooft couplings, one per gauge group, and the ratio of the two deforms continuously a $\mathbb{Z}_2$ orbifold of $\mathcal{N}=4$ into the $\mathcal{N}=2$ SCQCD. The dual string background of this theory is known to be $AdS_5\times S_5/\mathbb{Z}_2$, and some insight into the SCQCD dual background was obtained in \cite{Gadde2009} from this (singular) limit of the two couplings ratio. After localization, the $SU(N)\times SU(N)$ gauge theory is described by a quiver matrix model: the two Wilson loops, associated to each gauge group, are coupled trough their saddle point equation \cite{Rey2010,Zhou2009}. Although remarkable, the calculation presented in \cite{Passerini2011} may be difficult to extend to the quiver matrix model. We propose here a simpler method to study the saddle point equation \ref{equ_orig} at strong coupling that still retains the qualitative aspects of the problem. Keeping in mind a later application to the quiver problem, we will systematically compare our results with those derived in \cite{Passerini2011} in order to determine the accuracy of the approximation.

It can be shown by a rescaling of the eigenvalues that the large $\l$ limit amount to send the spectral parameters $x,y$ to infinity \cite{Rey2010}. We suggest to truncate the function $k$'s large $x$ expansion in the kernel and potential of \ref{equ_orig},
\begin{equation}\label{exp_k}
k(x)=2x\log|x|+2\g x+\sum_{k=0}^\infty{(-1)^k\dfrac{B_{2k+2}}{k+1}x^{-(2k+1)}},\quad x\to\pm\infty,
\end{equation}
where $B_k$ denote the Bernouilli numbers. The replacement of the kernel is justified for a large distance interaction between eigenvalues, i.e. when $|x-y|$ is large. When the eigenvalues are close to each others, the kernel $k$ is negligible compared to the Cauchy kernel, and the effect of the replacement should remain small. The approximation of the potential for $|x|$ large pertains to the Wilson loop computation: the dominant term is given by the largest eigenvalue which tends to infinity at strong coupling. Although convenient, this second approximation is not strictly necessary, and the solution for the full potential will be given in the concluding section.

We restrict ourselves to the first orders of the expansion \ref{exp_k}, for which the integral equation can be cast into the form\footnote{We use here the invariance of the equation under shifts of the kernel $k$ by a linear function which is a consequence of the density parity (see section 2 below).}
\begin{equation}\label{equ_sc}
\int_\G{\dfrac{\rho(y)dy}{x-y}}+\a\int_\G{\ksc(x-y)\rho(y)dy}=\b x+\a\ksc(x),\quad \ksc(x)=x\log|x|-x.
\end{equation}
This simplified problem can be solved exactly, providing an explicit expression for the density. Although less precise, this expression may appear more convenient than the one found in \cite{Passerini2011}. It will be used to compute the Wilson loop with the formula \ref{def_Wilson}. It can also be employed to derive the planar free energy which is the effective action taken at the saddle point,
\begin{equation}\label{Seff}
\mathcal{F}_0=-S_\text{eff}[\rho]=-\int_\G{\rho(x)dx\left(\dfrac{8\pi^2}{\l}x^2-2L(x)\right)}+\int_\G{\rho(x)dx\int_\G{\rho(y)dy\left(\log|x-y|-L(x-y)\right)}}
\end{equation}
where $L$ is the logarithm of the G-Barnes functions, $L'=k$ and $L(0)=0$.

To have a hint on the effect of higher terms in the kernel truncation, we consider two different orders of approximation. At the first order, we take
\begin{equation}\label{approx_1}
k(x)\simeq 2\ksc(x)+2(\g+1) x,\quad \a_1=-2,\quad \b_1=\dfrac{8\pi^2}{\l}.
\end{equation}
As a refined approximation, we also take into account the $1/x$ term of the expansion \ref{exp_k} which is of the same order than the Cauchy kernel,
\begin{equation}\label{approx_2}
k(x)\simeq 2\ksc(x)+2(\g+1) x+\dfrac{B_2}{x},\quad \a_2=-\dfrac{12}{5},\quad \b_2=\dfrac{48}{5}\dfrac{\pi^2}{\l}.
\end{equation}
The effect of the subleading term $B_2/x$ in the potential, i.e. in the RHS of \ref{equ_orig}, will be neglected here as it remains subdominant compared to all other members. In the following, we use the notation $\a=-a^2$ (with $a>0$) and $\b=4\pi^2 a^2/\l$. The two approximations \ref{approx_1} and \ref{approx_2} will be referred to by the indices $1$ and $2$ respectively.

In this setting, we keep the linear term $\b x$ that comes from the Gaussian matrix potential and confines the eigenvalues. Consequently, the support of the density remains compact and the Wilson loop is finite. This method keeps all the general characteristics of the solution observed in \cite{Passerini2011}. In particular, we recover the logarithmic dependence in $\l$ of the support endpoints. The Wilson loop we computed exhibits a similar behavior than \ref{Wilson_PZ}, with the approximate value $3.056$ of the cubic exponent (using the refined approximation \ref{approx_2}). It is argued that the exact exponent appears as a limit where all the terms of the expansion \ref{exp_k} are considered, and that this can be encoded in an effective value for the variable $a$, $a_\infty=\pi/2$.

From the solution of the approximate integral equation \ref{equ_sc}, we deduced the first orders of the planar free energy at large $\l$,
\begin{equation}\label{free_NRJ}
\CF_0\simeq \CF_0(\l=\infty)-\dfrac{16\pi^2}{a^2\l}+\dfrac{16\pi^4}{a^2}\dfrac{(\log\l)^2}{\l^2}+O\left(\dfrac{\log\l}{\l^2}\right)
\end{equation}
where $\CF_0(\l=\infty)$ is an integration constant we determined. This is one of the main results of this note, together with the Wilson loop expression \ref{Wilson_JEB}. The form of the $\l$-dependence in \ref{free_NRJ} does not involve $a$, it is believed to be identical to the exact result that would be obtained from the full density solving the initial problem \ref{equ_orig}.

\section{Solution of the integral equation}
To solve the simplified integral equation \ref{equ_sc}, we notice that, like the original equation \ref{equ_orig}, it is invariant under the sign inversion of the eigenvalues $x\to-x$. As a result, we look for an even density with a symmetric support $\G$. Since the eigenvalues are real, this support lies on the real axis, and we concentrate on a single cut solution for which $\G=[-M,M]$. The density $\rho$ has to vanish at the endpoints of its support, and is normalized to one. Due to the density parity, the first moment is vanishing which implements the $SU(N)$ constraint.

The first step to solve \ref{equ_sc} is to transform the Cauchy kernel into the $\ksc$ kernel. It is done using an integration by parts,
\begin{equation}\label{main_id}
\int_\G{\dfrac{\rho(y)dy}{x-y}}=\int_\G{\ksc(x-y)\rho''(y)dy}-\rho'(M)\left[\ksc(x+M)+\ksc(x-M)\right].
\end{equation}
We will see below that $\rho'(M)$ is actually infinite, and this identity only makes sense after a proper regularization. We will elaborate on this subtle point later, and for now we formally use \ref{main_id} to rewrite the saddle point equation as an integral equation with kernel $\ksc$ over the auxiliary density $\vphi=\rho''+\a\rho$,
\begin{equation}
\int_\G{\ksc(x-y)\vphi(y) dy}=\b x+\a\ksc(x)+\rho'(M)\left[\ksc(x+M)+\ksc(x-M)\right].
\end{equation}
In doing so, the potential acquires a boundary term. Since $\vphi$ is even, this equation is equivalent to its derivative,\footnote{An integral equation with the kernel $\ksc$ and a constant potential cannot have an even solution, so that the integration constant is automatically zero.}
\begin{equation}\label{equ1}
\int_\G{\log|x-y|\vphi(y) dy}=\b+\a\log|x|+\rho'(M)\log\left(M^2-x^2\right).
\end{equation}
This weakly singular integral equation is linear in $\vphi$, and the full solution is a sum of three components associated to the three members of the potential. For each of these solutions, that we denote $\vphi_i$, we have to solve an inhomogeneous differential equation to recover the associated density $\rho_i$,
\begin{equation}\label{equ_diff_rho}
\rho_i''(x)+\a\rho_i(x)=\vphi_i(x).
\end{equation}
The integration constants are fixed by imposing the parity of the density and the vanishing condition at the support endpoints. Below, we discuss each solution $\rho_i$ individually and it is convenient to fix the integration constants independently on each $\rho_i$.

\subsection{Solution for the boundary term}
We first look at the solution for the boundary term, i.e. the last term in the RHS of \ref{equ1}. This solution $\vphi_I$ is simply a sum of two $\delta$ functions supported on the endpoints $x=\pm M$. The differential equation \ref{equ_diff_rho} for the density can be solved using a Fourier transform, and we get
\begin{equation}\label{expr_rhoI}
\rho_I(x)=-\dfrac{\rho'(M)}{2a}\left(e^{-a|x+M|}+e^{-a|x-M|}-2e^{-aM}\cosh ax\right).
\end{equation}
This density is vanishing on the whole support $\G$ and cannot contribute to any regular integral. Consequently, it has a zero norm and gives no Wilson loop contribution. As we will see in the subsection 2.3, the main role of $\rho_I$ is to cancel the divergent terms that appear due to the integration by parts we employ. The derivative of $\rho_I$ is supported on the two endpoints of $\G$ where it takes the values
\begin{equation}
\rho'_I(\pm M)=\pm \hf \rho'(M).
\end{equation}
The unknown quantity $\rho'(M)$ in the expression \ref{expr_rhoI} of $\rho_I$ will thus be determined once we know the two other components $\rho_{II}$ and $\rho_{III}$ of the density $\rho$,
\begin{equation}\label{rel_deriv}
\hf\rho'(M)=\rho'_{II}(M)+\rho'_{III}(M).
\end{equation}

\subsection{Solution at $\b=0$}
The density $\vphi_{II}$ solves the integral equation \ref{equ1} for the logarithmic term in the potential. It is given by a delta function, now supported at $x=0$, which leads to the density
\begin{equation}
\rho_{II}(x)=\dfrac{a}{2}\dfrac{\sinh a(M-|x|)}{\cosh aM}
\end{equation}
after solving the differential equation. At infinite coupling $\b=0$, $\rho_{II}$ is the full solution, up to the boundary term $\rho_I$. As already observed in \cite{Passerini2011}, the unit norm condition on $\rho_{II}$ leads to a non-compact support, 
\begin{equation}\label{NII}
\int_\G{\rho_{II}(x)dx}=1-\dfrac1{\cosh(aM)}=1\implies M\to\infty.
\end{equation}
This is due to the vanishing of the Gaussian potential that confines the eigenvalues. In the next subsection, we introduce the density $\rho_{III}$ associated to this confining potential, i.e. to the constant term in the RHS of Equ. \ref{equ1}. This density $\rho_{III}$ contributes to the norm of $\rho$, and $M$ remains finite. 

Since at $\b=0$ the support $\G=\mathbb{R}$ is non-compact, the initial equation \ref{equ_orig} with complete kernel can be solved by Fourier transform \cite{Passerini2011}. It gives us the opportunity to compare our solution for the approximated kernel
\begin{equation}\label{rho2_inf}
\rho_{II}(x)=\dfrac{a}{2}e^{-a|x|}\quad \text{at  }M\to\infty,
\end{equation}
to the exact solution of the problem,
\begin{equation}\label{rhoPZ_inf}
\rho_\infty(x)=\dfrac1{2\cosh(\pi x/2)}= e^{-\pi |x|/2}+O\left(e^{-3\pi|x|/2}\right),
\end{equation}
in the strong coupling limit $|x|\to\infty$. We note that the coefficients inside the leading exponentials are different,
\begin{equation}\label{num_val}
\pi/2\simeq 1,571,\quad a_1=\sqrt{2}\simeq1,414,\quad a_2=2\sqrt{3/5}\simeq 1,549,
\end{equation}
but the numerical discrepancy is rather small (1.4\% for the finest approximation \ref{approx_2}). This discrepancy can be easily explained in the Fourier transform framework. Our strong coupling approximations of the kernel \ref{approx_1} and \ref{approx_2} amount to expand the Fourier transform $\hat k(\o)$ of $k$ at small $\o$.  Then, the two Fourier transformed densities agree at the first orders $\hat\rho_\infty(\o)\simeq\hat\rho_{II}(\o)$ but this is not enough to ensure the correct asymptotic in the $x$-space. However, given the proximity of the numerical value in \ref{num_val}, we expect a good approximation for the cubic exponent in the Wilson loop expression \ref{Wilson_PZ}, with an error of only a few percents.

We further notice that the coefficient $a$ in \ref{num_val} seems to converge toward the effective value $a_\infty=\pi/2$ as we refine the approximation for the kernel. To justify this fact, we may consider the next orders in the expansion \ref{exp_k} of the function $k$. After integration by parts, it leads to define an auxiliary density involving higher derivatives in $\rho$. Neglecting boundary terms, this density $\vphi$ still satisfies the equation \ref{equ1} with $\ksc$ kernel and a similar potential. When taking into account the whole expansion \ref{exp_k} for $k$, the auxiliary density writes
\begin{equation}\label{def_vphi}
\vphi=\rho''-2\rho-\sum_{m=0}^\infty{\dfrac{(-1)^mB_{2m+2}}{(2m)!(m+1)}\rho^{(2m+2)}}.
\end{equation}
The coefficient $a$ is related to the homogeneous solution of the differential equation and can be obtained as a root of the characteristic equation. The characteristic equation for the infinite differential system \ref{def_vphi} takes the form of an infinite series. Remarkably, the summation can be performed and we are led to the very simple equation,
\begin{equation}
2\sin^2(a/2)=1.
\end{equation}
The solutions are $a_\infty=\pi/2$ modulo $\pi$. Since the density should vanish at $|x|\to\infty$, we have to select only the decreasing exponentials, and we recover the dominant contribution of \ref{rhoPZ_inf}, as well as all the exponentially small corrections.

\subsection{Complete solution}
We now analyze the last part of the density, $\rho_{III}$, which is associated to the constant term $\b$ in the potential of \ref{equ1}. The integral equation for the auxiliary density $\vphi_{III}$ does not depend on $\a$, and we first take $\a=0$. Then, the integral equation \ref{equ_sc} on $\rho$ reduces to the $\mathcal{N}=4$ saddle point equation which is solved by the Wigner semi-circle law,
\begin{equation}
\rho(x)=\dfrac{\b}{\pi} R(x),\quad R(x)=\sqrt{M^2-x^2}.
\end{equation}
At $\a=0$, $\vphi_{III}$ is equal to the second derivative of the Wigner density, up to a boundary term $\vphiB_{III}$,
\begin{equation}\label{ansatz}
\vphi_{III}(x)=-\dfrac{\b M^2}{\pi R(x)^3}+\vphiB_{III}(x).
\end{equation}
Pluging this ansatz into the integral equation \ref{equ1}, we indeed find a solution for $\vphiB_{III}$ as a sum of two delta functions supported on the endpoints,
\begin{equation}
\vphiB_{III}(x)=\dfrac{\b M}{\pi R(M)}(\d(x+M)+\d(x-M)).
\end{equation}
This expression, obtained after integration by parts, contains the infinite term $1/R(M)$. There are several ways to regularize this computation, the simplest is to replace the Wigner solution $R(x)$ by the function
\begin{equation}
R_\e(x)=\sqrt{(M+\e)^2-x^2}.
\end{equation}
The corresponding density provides the correct solution up to $O(\e)$ terms, and $R_\e$ is non-vanishing at $x=\pm M$. To avoid unnecessary complicated expressions, we will not explicitly write down this regularization here, and we keep the notation $1/R(M)$ to imply $O(1/\sqrt{\e})$ divergent terms that have to be canceled. This cancellation will be achieved using the density $\rho_I$, which shows that the infinite terms appearing here are just an artifact of the integration by parts.

Returning to $\a\neq 0$, we have to solve the differential equation \ref{equ_diff_rho} on $\rho_{III}$ with the second member $\vphi_{III}$ given by Equ. \ref{ansatz}. We get
\begin{equation}\label{expr_rhoIII}
\rho_{III}(x)=-\dfrac{\b M^2}{\pi a}\left[ \int_0^x{\dfrac{\sinh a(x-y)}{R(y)^3}dy}-\dfrac{\cosh ax}{\cosh aM}\int_0^M{\dfrac{\sinh a(M-y)}{R(y)^3}dy}\right]+\rho_{III}^{(B)}(x),
\end{equation}
and a boundary term which is similar to the expression \ref{expr_rhoI} of $\rho_I$,
\begin{equation}
\rho_{III}^{(B)}(x)=-\dfrac{\b M}{2a\pi R(M)}\left(e^{-a|x+M|}+e^{-a|x-M|}-2e^{-aM}\cosh ax\right).
\end{equation}
We used here the homogeneous solution to impose the vanishing of the density at $x=\pm M$. The expression \ref{expr_rhoIII} for $\rho_{III}$ is even and reduces to the Wigner solution at $\a=0$.

Once $\rho_{II}$ and $\rho_{III}$ are known, we can determine the constant $\rho'(M)$ involved in the expression of $\rho_I$ using the relation \ref{rel_deriv}. The derivative of $\rho_{III}$ at the endpoints contains a $1/R(M)$ singularity,\footnote{This calculation is a bit delicate and we have to use the formal formula
\begin{equation}
\int_0^M{\dfrac{\cosh{ay}}{R(y)^3}dy}=\dfrac{\cosh(aM)}{MR(M)}-\dfrac{\pi a}{2M}I_1(aM).
\end{equation}}
\begin{equation}
\rho_{III}'(M)=-\dfrac{\b M}{2\pi R(M)}+\hf a\b M\dfrac{I_1(aM)}{\cosh aM},
\end{equation}
where $I_1$ denotes the Bessel function of the first kind. From this result and the derivative of $\rho_{II}$ at $x=M$,
\begin{equation}
\rho'_{II}(M)=-\dfrac{a^2}{2\cosh aM},
\end{equation}
we deduce the expression of $\rho_I$. As required, the divergent terms in $\rho_I$ and $\rho_{III}^{(B)}$ cancel each others, and we are led with a finite expression for the density. This final density can be written as a sum of three components, $\rho=\tilde{\rho}_I+\rho_{II}+\tilde{\rho}_{III}$ with
\begin{align}
\begin{split}\label{sol_density}
\tilde{\rho}_I(x)&=\dfrac{a-\b MI_1(aM)}{2\cosh aM}\left(e^{-a|x+M|}+e^{-a|x-M|}-2e^{-aM}\cosh ax\right),\\
\rho_{II}(x)&=\dfrac{a}{2}\dfrac{\sinh a(M-|x|)}{\cosh aM},\\
\tilde{\rho}_{III}(x)&=-\dfrac{\b M^2}{\pi a}\left[ \int_0^x{\dfrac{\sinh a(x-y)}{R(y)^3}dy}-\dfrac{\cosh ax}{\cosh aM}\int_0^M{\dfrac{\sinh a(M-y)}{R(y)^3}dy}\right].
\end{split}
\end{align}\\

We still have to determine the dependence in $\l$ of the support $\G$, which is usually achieved by imposing the unit norm condition on $\rho$. The norm of $\rho$ have contributions only from $\rho_{II}$ (Equ. \ref{NII}) and $\tilde{\rho}_{III}$,
\begin{equation}\label{NIII}
\int_\G{\tilde{\rho}_{III}(x)dx}=\dfrac{\b MI_1(aM)}{a\cosh aM}.
\end{equation}
Combining these two results, we deduce the relation
\begin{equation}\label{norm}
 aMI_1(aM)=\dfrac{\l}{4\pi^2}. 
\end{equation}
When $\l$ is infinite, $M$ should also be infinite, as seen in \ref{NII}. Using the asymptotic of the Bessel function, the previous relation simplifies into
\begin{equation}\label{expr_M}
C\sqrt{M}e^{aM}=\l\implies M(\l)\simeq \dfrac1a\log\l-\dfrac1{2a}\log\log\l+\cdots
\end{equation}
where $C$ is a constant factor. We recover the logarithmic dependence of the endpoint $M(\l)$ found in \cite{Passerini2011} and specific to SCQCD. More precisely, the relation between $M$ and $\l$ is identical to the Equ. (4.7) of \cite{Passerini2011}, provided we use the effective value $a_\infty=\pi/2$. The constant $C$ should be compared to the value $14.60$ obtained in \cite{Passerini2011}, and to the outcome of the numerical simulation $C_\text{INT}\simeq15.02$ (appendix B of \cite{Passerini2011})
\begin{equation}
C=(2\pi)^{3/2}\sqrt{a},\quad C_1\simeq18.73,\quad C_2\simeq19.60,\quad C_\infty\simeq19.74,
\end{equation}
where the index $\infty$ refers to the use of the effective value $a_\infty$. We note that our method gives the correct order of magnitude for $C$, but does not provide a very accurate approximation for this constant.

\subsection{Wilson loop}
The Wilson loop is computed using the formula \ref{def_Wilson} with the solution \ref{sol_density} of the density. There are two contributions,
\begin{align}
\begin{split}
&W_{II}(\l)=\dfrac{a^2}{4\pi^2-a^2}\left(\dfrac{\cosh 2\pi M}{\cosh aM}-1\right),\\
&W_{III}(\l)=\dfrac{\b M}{4\pi^2-a^2}\left(2\pi I_1(2\pi M)-a\dfrac{\cosh2\pi M}{\cosh aM}I_1(aM)\right).
\end{split}
\end{align}
These expressions are exact in the approximation \ref{equ_sc} of the saddle point equation. Interestingly, the dominant term at strong coupling, $W_{III}$, exhibits a compact expression as a Bessel function of $M$. This is reminiscent of $\mathcal{N}=4$ super Yang-Mills \cite{Erickson2000,Drukker2000}, and one could speculate about a Laguerre polynomial formulation for the $1/N$ corrections \cite{Drukker2000}. At the first order, the Bessel functions can be approximated by their asymptotic,\footnote{Here the coefficient $r$ differs roughly by a factor $2$ from the result $r\simeq2.18$ derived in \cite{Passerini2011},
\begin{equation}
r_1\simeq1.11,\quad r_2\simeq1.29,\quad r_\infty\simeq 1.32\ .
\end{equation}}
\begin{equation}
W(\l)\simeq r\dfrac{\sqrt{M}}{\l} e^{2\pi M},\quad\text{with}\quad r= \dfrac{(2\pi)^{3/2} a^2(\sqrt{2\pi}-\sqrt{a})}{4\pi^2-a^2}.
\end{equation}
The dependence of the Wilson loop in $M$ remains exponential, and the power growth of $W(\l)$ is due to the unusual logarithmic behavior \ref{expr_M} of endpoint $M(\l)$,
\begin{equation}\label{Wilson_JEB}
W(\l)\simeq A \left(\dfrac{\l}{\sqrt{\log \l}}\right)^{\frac{2\pi}{a}-1},\quad A=\dfrac{(2\pi)^{3/2-3\pi/a} a^{3/2}(\sqrt{2\pi}-\sqrt{a})}{4\pi^2-a^2}.
\end{equation}
Using the refined approximation \ref{approx_2}, we get the exponent $3.056$, which differs only by 2\% from the cubic exponent found by Passerini and Zarembo. This cubic exponent arises at the effective value $a_\infty=\pi/2$ which takes into account the neglected higher order terms in the kernel approximation. Our approximation becomes less accurate for the constant factor $A$,
\begin{equation}
A_1\simeq4.46\times 10^{-6} ,\quad A_2\simeq1.44\times 10^{-5},\quad A_\infty\simeq 1.71\times 10^{-5},
\end{equation}
which differs roughly by a factor five from the value $A\simeq9.47\times 10^{-5}$ derived in \cite{Passerini2011}.

\subsection{Free energy}
To find the planar free energy, we should inject the expression of the density into the effective action \ref{Seff}. But it is actually easier to first consider the $\l$-derivative which is simply the second moment of the density,
\begin{equation}\label{deriv_CF}
\p_\l\mathcal{F}_0[\rho]=\dfrac{8\pi^2}{\l^2}\int_\G{x^2\rho(x)dx}.
\end{equation}
This relation can be obtained by taking the derivative of $\CF_0[\rho]$ seen as a functional of $\rho$, and then use the solution $\rho=\rho^*$ of the saddle point equation. It can also be checked explicitly that
\begin{equation}\label{CFdl}
\dfrac{d}{d\l}\left(\CF_0[\rho^*]\right)=(\p_\l\CF_0)[\rho^*]-2\int_\G{\p_\l\rho^*(x) V_\text{eff}(x)dx},
\end{equation}
where
\begin{equation}
V_\text{eff}(x)=\dfrac{4\pi^2}{\l}x^2-L(x)-\int_\G{\rho(y)dy\left[\log|x-y|-L(x-y)\right]}
\end{equation}
is the effective potential associated to the integral equation \ref{equ_orig}. Note that the $\l$-dependence of the support gives no contribution since $\rho(M)=0$. When $\rho$ satisfies the saddle point equation, the effective potential is a constant and can be removed from the integration in the RHS of \ref{CFdl}. This term becomes proportional to the derivative of the density norm, which vanishes since $\rho$ has norm one.

The second moment of the density remains finite at strong coupling, and the leading order of the free energy $\l$-derivative \ref{deriv_CF} can be computed using the infinite density profile $\rho_\infty$ given in \ref{rhoPZ_inf},
\begin{equation}\label{ex_deriv_CF}
\p_\l\mathcal{F}_0=\dfrac{8\pi^2}{\l^2}\int_\mathbb{R}{x^2\rho_\infty(x)dx}+o\left(\dfrac1{\l^2}\right)=\dfrac{8\pi^2}{\l^2}+o\left(\dfrac1{\l^2}\right).
\end{equation}
This expression can be integrated to obtain $\mathcal{F}_0$,
\begin{equation}\label{CF_res1}
\mathcal{F}_0=\CF_0(\l=\infty)-\dfrac{8\pi^2}{\l}+o\left(\dfrac1{\l}\right),
\end{equation}
where the integration constant is the free energy at infinite coupling. To find this constant, we notice that the insertion of the effective potential in the expression \ref{Seff} of the planar free energy leads to a huge simplification,
\begin{equation}\label{CF}
\CF_0(\l)=\int_{\G}{\rho^*(x)\log|x|dx}-\dfrac{4\pi^2}{\l}\int_\G{x^2\rho^*(x)dx}.
\end{equation}
At infinite coupling, the second term vanishes and we can use again the density $\rho_\infty$ to compute the integration constant,
\begin{equation}\label{CF_res2}
\CF_0(\l=\infty)=\int_{\mathbb{R}}{\rho_\infty(x)\log|x|dx}=2\log\left(\dfrac{2\G[3/4]}{\G[1/4]}\right)\simeq-0.7832.
\end{equation}

The first and second orders of the planar free energy at strong coupling, resp. \ref{CF_res2} and \ref{CF_res1}, were computed exactly and no approximation of the integral equation \ref{equ_orig} has been used. To go beyond these first orders, we need to take into account the correction to the density due to the confining potential of the integral equation. This is exactly the role played by $\rho_{III}$ for the strong coupling approximation of the integral equation. Thus, the density \ref{sol_density} determined previously should give us a hint on the form of the subleading term for the free energy, but at the price of a distortion on the numerical coefficients. In particular, $\rho_\infty$ will be replaced by $\rho_{II}$.\footnote{Computing the free energy, the potential approximation at large $|x|$ is not fully justified. However, we shall see in the next section that it does not modify the form of the subleading term.}

The relations \ref{deriv_CF} and \ref{CF} are exact provided that $\rho$ solves the initial saddle point equation \ref{equ_orig}. When employed with the density solving the equation with approximate kernel \ref{equ_sc}, it assumes a similar strong coupling approximation for the expression \ref{Seff} of $\CF_0[\rho]$. The calculation of the second moment for the density \ref{sol_density} is relatively straightforward, and using the normalization condition \ref{norm} the free energy derivative simplifies into
\begin{equation}\label{d_F0}
\p_\l\mathcal{F}_0=\dfrac{16\pi^2}{a^2\l^2}\left[1-\dfrac{2\pi^2 a^2 M^2}{\l}\right]\simeq\dfrac{16\pi^2}{a^2}\dfrac1{\l^2}-\dfrac{32\pi^4}{a^2}\dfrac{(\log\l)^2}{\l^3}+\cdots.
\end{equation}
Using the relation \ref{expr_M} between $M$ and $\l$, the expression \ref{d_F0} can be integrated to obtain $\CF_0$,
\begin{align}\label{CF0}
\begin{split}
\CF_0&\simeq \CF_0(\l=\infty)-\dfrac{16\pi^2}{a^2}\dfrac1{\l}+\dfrac{2\pi}{a^2}(1+aM)e^{-2aM}\\
&\simeq \CF_0(\l=\infty)-\dfrac{16\pi^2}{a^2}\dfrac1{\l}+\dfrac{16\pi^4}{a^2}\dfrac{(\log\l)^2}{\l^2}+O\left(\dfrac{\log\l}{\l^2}\right).
\end{split}
\end{align}
As expected, the first and second order terms have the same form than the exact result, with different numerical coefficients. The next order contains a logarithmic non-analyticity that would be interesting to interpret in the gauge theory context. To test the distortion of the numerical coefficients, we computed the integration constant for the approximate density $\rho_{II}$,
\begin{equation}
\CF_0(\l=\infty)=\int_{\mathbb{R}}{\rho_{II}(x)\log|x|dx}=-\g-\log a_2\simeq -1.015,
\end{equation}
the numerical discrepancy between the exact result and the approximation is about 30\%. The subleading coefficients differ by 17\% in the approximation \ref{approx_2}. 


\section{Discussion}
In this note, we discussed an alternative method to study the saddle point equation for the Wilson loop in SCQCD. We used a strong coupling approximation for the kernel and potential of the saddle point equation, and derived an explicit solution for the density of the simplified problem. From this solution, we computed the Wilson loop dominant contribution at strong coupling and compared it to the expression derived in \cite{Passerini2011}. We found a good numerical approximation for the exponents characterizing the dependence in $\l$ and $\log\l$. The exact exponents can be obtained from an effective parameter $a_\infty$ that takes into account the effect of neglected terms in the kernel strong coupling approximation. The simplified problem exhibit the same qualitative behavior than the original one, and our approach should be suitable to simplify the quiver matrix models when at least one of the gauge couplings is strong. A similar technique can also be employed to study the expressions of the 't Hooft loops recently derived for $\mathcal{N}=2$ gauge theories from localization \cite{Gomis2011}.

We computed the leading orders of the free energy, the subdominant term exhibits an intriguing $1/\l$ behavior that differs from the logarithmic result found in $\mathcal{N}=4$ super Yang-Mills theory,
\begin{equation}
\mathcal{F}_0^{(\mathcal{N}=4)}=\hf\log\l-\dfrac14-\log(4\pi).
\end{equation}
Reintroducing the variables $N$ and $g$, and taking into account the $1/N^2$ factor in front of the planar free energy, the SCQCD free energy behaves at strong coupling as $\CF\simeq 1/N^2+1/g^2N^3$ (we omitted the constants). This information could help to select the correct background of the dual string theory.

Until now, we mostly focused on the kernel approximation. To investigate the approximation of the potential we consider
\begin{equation}
\int_\G{\dfrac{\rho(y)dy}{x-y}}+\a\int_\G{\ksc(x-y)\rho(y)dy}=\b x+\dfrac{\a}{2}\left(k(x)-2(\g+1)x\right)
\end{equation}
where we kept the full potential of \ref{equ_orig}. Since the integral equation is linear in the density, we only need to introduce an additional contribution $\rho_{IV}$ to the total density $\rho$ to take into account the higher order terms. Using the asymptotic expansion \ref{exp_k}, it is not hard to see that 
\begin{equation}
\rho_{IV}(x)=-\dfrac{a^2}{2}\sum_{k=0}^\infty{\dfrac{B_{2k+2}}{k+1} \dfrac1{(2k)!}\sum_{m=0}^k{\a^m (-1)^{k-m}\d^{(2k-2m)}(x)}}+\hf f(a)\rho_{II}(x).
\end{equation}
with
\begin{equation}
f(a)=\sum_{k=0}^\infty{\dfrac{B_{2k+2}}{k+1} \dfrac{(-1)^k}{(2k)!}a^{2k+2}}=\dfrac{a^2}{2\sin^2(a/2)}-2
\end{equation}
solves the problem. This density contributes to the normalization of $\rho$, which modifies the constant $C$ in the relation \ref{expr_M} giving $M(\l)$,
\begin{equation}
C=4(2\pi)^{3/2}a^{-3/2}\sin^2(a/2),\quad C_1\simeq15.81,\quad C_2\simeq15.98,\quad C_\infty=16.
\end{equation}
These figures are closer to the value $C=14.60$ found in \cite{Passerini2011}. The contribution $W_{IV}$ to the Wilson loop remains subdominant, but the modification of the relation $M(\l)$ affects the prefactor $A$ in the expression of $W(\l)$,
\begin{equation}
A_1\simeq9.48\times 10^{-6} ,\quad A_2\simeq3.30\times 10^{-5},\quad A_\infty\simeq 3.95\times 10^{-5},
\end{equation}
which now only differs roughly by a factor two from the numerical value given in \cite{Passerini2011}. The density $\rho_{IV}$ also contributes to the derivative of the free energy, and we get in the approximation \ref{approx_2},
\begin{equation}
\p_\l\mathcal{F}_0=\dfrac{8\pi^2}{\l^2}\left[1-\dfrac{4\pi^2 M^2}{\l}\right]\simeq \dfrac{8\pi^2}{\l^2}-\dfrac{32\pi^4}{a_2^2}\dfrac{(\log\l)^2}{\l^3}+\cdots
\end{equation}
We notice that the dependence in $\l$ remains the same when we consider higher corrections for the potential, but the overall constant factors get modified. In this way, we reach a better agreement when comparing to the results of \cite{Passerini2011}. But these numerical factors remain rather sensitive to the approximation, and we cannot rely on our approach for their derivation. On the other hand, the $\l$-dependence is quite strong and can be trusted.


So far, we mainly concentrate on the technical aspects of the problem, and these results still have to be understood within the gauge theory framework. In particular, it would be interesting to see how the logarithmic non-analyticities appear at strong coupling. In this context, we cannot employ the kernel approximation, and should study the original equation \ref{equ_orig}, which solution interpolates between strong and weak coupling regimes. The kernel function $k$ has poles on the imaginary axis at $x\in i\mathbb{Z}$. To each pole is associated a copy of the eigenvalue density support where the resolvent of the corresponding Riemann-Hilbert problem has a branch cut. In the weak coupling limit, $x$ is small and the branch cuts associated to the imaginary poles are send to infinity. Only the branch cut $\G$ lying on the real axis remains. On the other hand, at strong coupling these branch cuts condense on the real axis, leading to the logarithmic behavior for the kernel. One may wonder if something similar happen within the gauge theory.


\section*{Acknowledgements}
I would like to acknowledge Chuan-Tsung Chan, K. Hosomichi, H. Irae, Y. Matsuo and Y. Zhou for valuable discussion, and more particularly S.-J. Rey, T. Suyama and K. Zarembo for helpful comments on previous versions of this note. It is also a pleasure to thanks the warm hospitality of NCTS and NTU in Taiwan where a part this work was completed. This work is partially supported by the National Research Foundation of Korea (KNRF) grant funded by the Korea government (MEST) 2005-0049409.

\bibliographystyle{unsrt}

\end{document}